\documentclass[onecolumn]{aa} 

\usepackage{graphicx}
\usepackage{url}
\usepackage{tabularx}
\usepackage{longtable}
\usepackage{hyperref}
\hypersetup{
    colorlinks=true,
    linkcolor=red,
    filecolor=magenta,      
    urlcolor=magenta,
    citecolor=blue,
}
\usepackage{txfonts}
\usepackage{rotating}
\usepackage{lscape}
\usepackage{multirow}
\usepackage{subcaption}
\usepackage{adjustbox}
\usepackage{amsmath}
\usepackage{multirow}
\usepackage{lscape}
\usepackage{longtable}

\begin{document} 

   \title{Can metric radio bursts be used as a diagnostics tool for interplanetary coronal mass ejections?}


   \author{J. Kandekar \inst{1}
   \and
   A. Kumari \inst{2,3}\fnmsep\thanks{Corresponding Author \email{anshu@prl.res.in}}
   }

   \institute{
    Department of Physics, Ahmednagar College, Station Road, Ahilyanagar-414001, Maharashtra, India
   \and
   Udaipur Solar Observatory, Physical Research Laboratory, Dewali, Badi Road, Udaipur-313 001, Rajasthan, India
   \and
  Department of Physics, University of Helsinki, P.O. Box 64, FI-00014, Helsinki, Finland \\
   \email{anshu.kumari@helsinki.fi}
}


 
  \abstract
   {}
   {Metric radio bursts are often said to be valuable diagnostic tools for studying the near-sun kinematics and energetics of the Interplanetary Coronal Mass Ejections (ICMEs). Radio observations also serve as an indirect tool to estimate the coronal magnetic fields. However, how these estimated coronal magnetic fields are related to the magnetic field strength in the ICME at 1 AU has rarely been explored. We aim to establish a relation between the coronal magnetic fields obtained from the radio observations very close to the Sun and the magnetic field measured at 1 AU when the ICME arrives at the Earth.}
   {We performed statistical analysis of all metric type II radio bursts in solar cycles 23 and 24, which were found to be associated with ICMEs. We estimated the coronal magnetic field associated with the corresponding CME near the Sun (middle corona) using a split-band radio technique and compared those with the magnetic fields recorded at 1 AU with in-situ observations.}
   {We found that the estimated magnetic fields near the Sun using radio techniques are not well correlated with the magnetic fields measured at 1 AU using in-situ observations. This could be due to the complex evolution of the magnetic field as it propagates through the heliosphere. } 
   {
   Our results suggest that while metric radio observations can serve as effective proxies for estimating magnetic fields near the Sun, they may not be as effective close to the Earth. At least, no linear relation could be established using metric radio emissions to estimate the magnetic fields at 1 AU with acceptable error margins.}

   \keywords{ Sun: coronal mass ejections (CMEs), Sun: -- activity, Sun: -- radio radiation, Sun: -- sunspots, Sun: -- corona }

   \maketitle
%

\section{Introduction}
\label{section:sec1}

The coronal magnetic field (B) serves as a fundamental factor governing the formation, evolution, and dynamics of solar corona structures \citep{Dulk1978}. 
Measuring these fields directly from observations, particularly in the chromosphere and corona, remains a significant challenge in solar physics 
\citep[][and the references therein]{White2002, Carley2020}. Various methods exist to infer coronal magnetic fields, encompassing techniques like coronal seismology \citep{Jess2016}, Zeeman splitting of spectral lines \citep{lin2004coronal}, and extrapolation methods \citep{Wiegelmann2017}, though constrained to specific regions of the solar atmosphere. Among these, including the shock-standoff technique \citep{Gopalswamy2011} and radio techniques, estimations from solar radio bursts 
, offer valuable insights into the coronal magnetic field, especially in the middle corona \citep{Carley2017}.

Solar eruptive phenomena, such as solar flares and coronal mass ejections (CMEs), significantly influence space weather dynamics. 
These events often trigger energetic electron acceleration, generating radio emissions like type II and type IV bursts observed in the metric and decameter-hectometric (DH) frequency range \citep{Dulk1980}. 
Taking advantage of these radio bursts enables the study of CMEs' early evolution and dynamics near the Sun (within 10 solar radii), which can help in estimating CME magnetic fields and assessing their geo-effectiveness. Since radio observations consist of solar, heliospheric, and ionospheric space weather phenomena,  radio techniques can offer a practical approach for probing the solar atmosphere's magnetic field \citep{White2004, Gopalswamy2006}. Combining diverse radio techniques with broadband spectroscopic solar radio imaging allows for deriving coronal magnetic fields in various solar regions \citep{ramesh2000radio, kumari2017b}. However, while metric type II radio bursts and their spectral characteristics, such as split-band features, have long served as robust tools for direct coronal magnetic field estimation \citep{Smerd1974, vrvsnak2008origin, 
2018JASTP.172...75M}, their correlation with magnetic fields observed at 1 AU during near-earth interplanetary coronal mass ejections (ICMEs) remains less explored.

Previously, researchers have investigated metric and DH radio bursts and their correlation with electron accelerations, coronal mass ejections (CMEs), and their variations throughout the solar cycle \citep{Kahler2019, 
Kumari2023b}. However, to the best of our knowledge, no prior study has explored how the characteristics of metric radio bursts can provide insights into space weather phenomena, such as ICMEs. In this study, we investigate the long-term datasets of solar radio bursts and ICMEs over solar cycles 23 and 24 to explore the feasibility of using radio techniques for coronal magnetic field estimation and their relationship with in-situ magnetic field observations at 1 AU during ICME events. 

This article is structured as follows: Section \ref{section:sec2} discusses the observational data utilized in this study along with detailed data analysis procedures. The findings of the present study are presented in Section \ref{section:sec3}. Section \ref{section:sec4} contains a discussion of the results and their implications, followed by a summary of the study.

\section{Observations and Data Analysis}
\label{section:sec2}

For this study, we identified 88 metric type IIs associated with the ICMEs of solar cycles 23 and 24 (1997-2019). The CME event list used here was obtained from the Coordinated Data Analysis Workshop \citep[CDAW;][]{Yashiro2004} database\footnote{\url{https://cdaw.gsfc.nasa.gov/}}, which is a manual catalog of the CMEs and their properties recorded by the Solar and Heliospheric Observatory's (SOHO) Large Angle and Spectrometric Coronagraph (LASCO) white-light coronagraph observations \citep{Brueckner1995}. This list has the CME properties \& parameters such as linear and second-order speed, kinetic energy, mass, angle of position, etc. Additionally, we used the ICME list\footnote{{\url{https://izw1.caltech.edu/ACE/ASC/DATA/level3/icmetable2.htm}}} provided by \cite{Richardson2010}, containing the date and time of plasma disturbances of ICME, duration, magnetic field strength, and speed, among other details, and associated with the date-time of CME when it reaches 1~AU. This list consists of ICME events since January 1996. Following the CME events associated with the ICME events, the radio data obtained from the Solar Electro-Optical Network \citep[SEON][]{} database\footnote{\url{https://www.ncei.noaa.gov/products/space-weather/legacy-data/solar-electro-optical-network}} was used for the data analysis. 
It consists of two more networks: 1) the Solar Observing Optical Network (SOON) and 2) the Radio Spectral Telescope Network (RSTN), containing four solar radio observatories across the globe. The radio database from the Radio Spectral Telescope Network (RSTN) was utilized; this data consists of the Solar Radio Spectrographs (SRS) over a wide band (25-180 MHz), wherein Radio Interference Measuring Sets (RIMS) is the instrument detecting radio emissions at eight different frequencies. For the missing data/type IIs in the SEON's RSTN database, eCallisto data\footnote{\url{https://www.e-callisto.org/links.html}} was referred. Furthermore, Hiraiso Radio Spectrograph (HiRAS) data \citep{1995CLRJ...42..111K} managed by the National Institute of Information and Communications Technology (NICT\footnote{\url{https://solarobs.nict.go.jp/}}) was also utilized for obtaining the spectra of type II bursts.
We used the OMNI 5-min data ( from NASA/GSFC's OMNIweb for the related ICMEs. The data was accessed through NASA's Coordinated Data Analysis Web (CDAWeb\footnote{\url{https://cdaweb.gsfc.nasa.gov/}\label{fn:6}}). We utilized the 5-min high-resolution flow speed (km/s) and the magnetic field (nT) data.

\begin{figure}[ht!]
        \includegraphics[width=7cm]{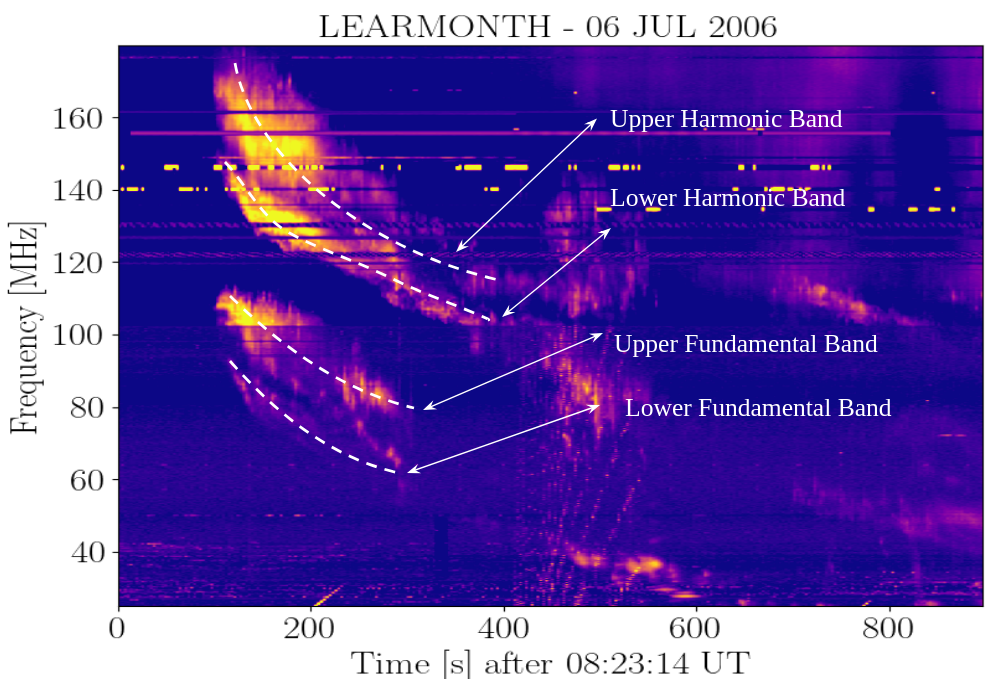}
        \includegraphics[width=8cm]{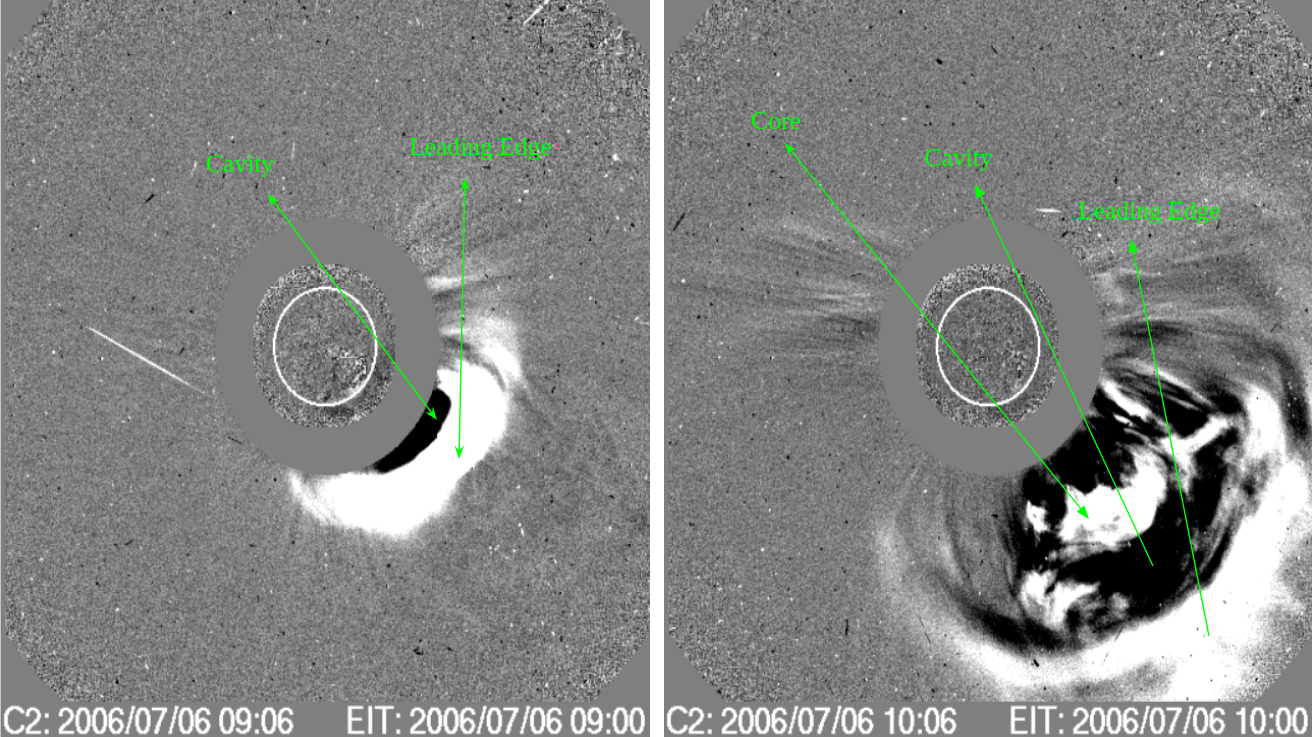}
    \caption{ \textit{The left panel}: Type II burst's dynamic spectrum recorded by Learmonth Observatory on 6 Jul, 2006, ranging from 180-25 MHz. The type II burst started at 08:24 UTC and lasted for 17 minutes. The white dotted lines represent the fundamental and harmonic bands.
    \textit{The right panel}: The three-part structure of the CME with a core, cavity, and leading edge recorded by LASCO C2 on 6 Jul, 2006 at 09:06 UTC and 10:06 UTC. The CME started at 08:54 UTC (Appendix \ref{tab:CME+typeII})}
    \label{fig:figure1}
\end{figure}

\begin{figure}[ht!]
    \centering
    \includegraphics[width=10cm]{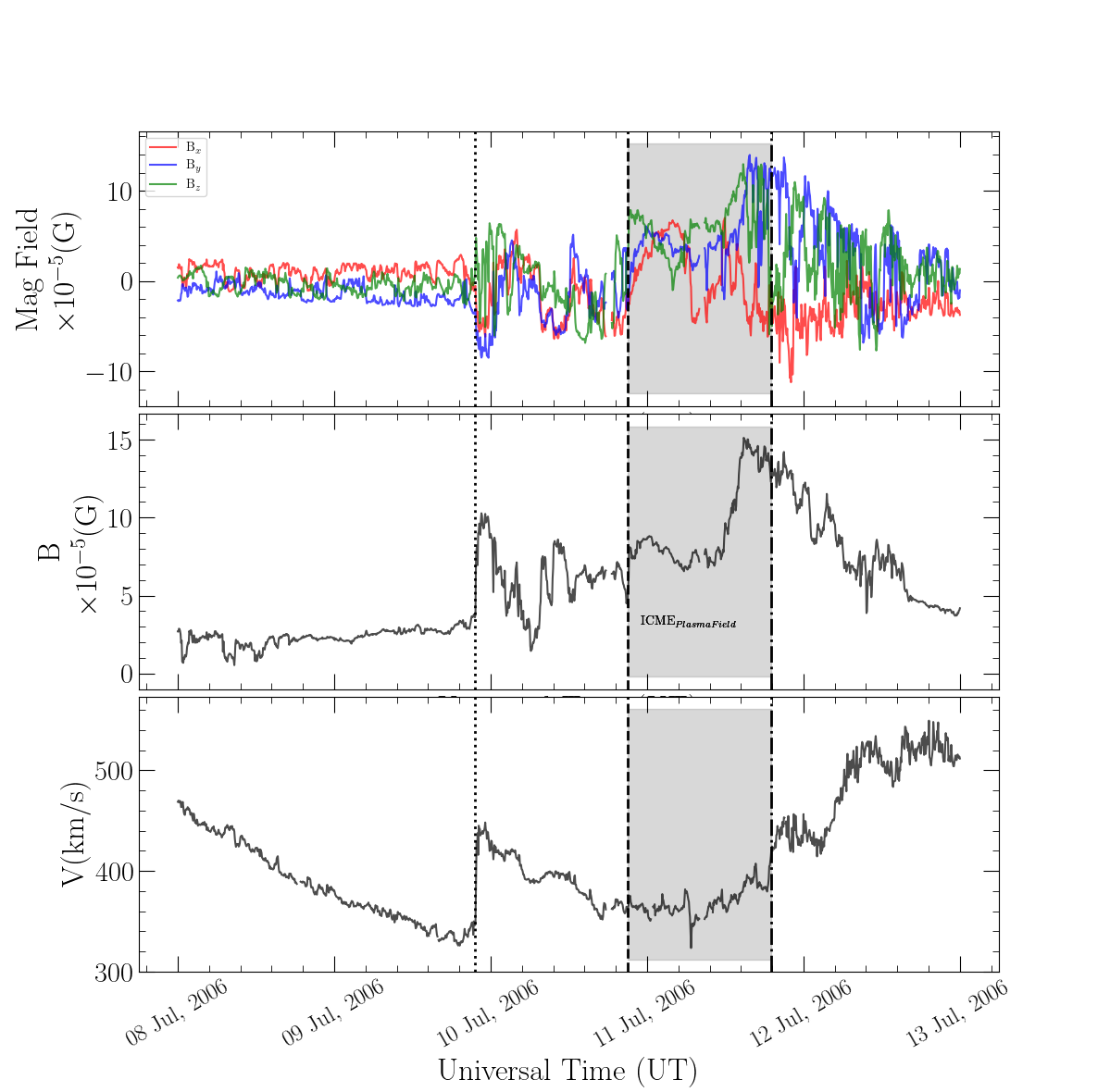}
    \caption{For the type II burst and CME that occurred on 6 Jul, 2006, the figure shows the associated plot of ICME parameters obtained from Omni data. The first subplot illustrates the Bx, By, and Bz vector components of the B field (G), the middle subplot shows the average strength of the B field vector(nT), and the third subplot shows the flow speed across the timeline of the event. The vertically dotted line shows the near-earth shock arrival at 21:36 UT on 9 Jul, 2006 and the grey-shaded region shows the start and end time of the disturbance in the plasma \citep{Richardson2010}, starting at 19:00 UT on 9 Jul, 2006 and ending at 21:00 UT on 11 Jul, 2006.}
    \label{fig:figure2}.
    
\end{figure}

Out of the 88 events identified, only 31 metric type IIs for the solar cycle of 23 (1996 -2008) and 24 (2008-2019) had split-band features identified in the solar dynamic spectra. 
Few of the type IIs had a start frequency $>$ 180 MHz (appendix \ref{tab:CME+typeII}). The 31 CME-ICME associated split-band events comprised a fundamental band and a harmonic band. As shown in Figure \ref{fig:figure1}, these events also exhibited a clear three-part structure \citep[core, cavity and leading edge,][]{2008ApJ...672.1221R}. 
For radio data analysis, the dynamic spectra were cleaned and radio frequency interference was removed. 
We used the 4-fold Newkirk density model \citep{newkirk1961solar} 
to convert the type II frequencies into radio heights. 
We used the following equation to estimate the magnetic field strength of 31 split-bands (CME-ICME associated events) \citep{Smerd1974, 
Vrsnak2002}
\begin{equation}
\label{eq:eqq}
    B =  5.1 \times 10^{-5} \times F_{LB}\times v_A ~ ~~~~~~~~[G]
\end{equation} 
where B is the magnetic field strength derived from the type II bursts in gauss, F$_{LB}$(MHz) is the lower band frequency of the "undisturbed Corona" \citep{Vrsnak2002
}, and v$_A$ (km/s) is the  Alfvén speed, the magnetic fields obtained were compared with the magnetic fields recorded at 1 AU. A master list of solar cycle 23 and 24 events was created and divided into two tables/lists. Table \ref{tab:CME+typeII} for type IIs and CME events consisting of date, start time (UT), end time (UT), duration (min), start \& end frequency (MHz), start time of CME (UT), CME's width (deg), and CME speed (km/s) obtained from the CDAW database. Table \ref{tab:ICME} of ICME events containing date, start time(UT), start and end time of the disturbances in plasma along with its start-end date, ICME Plasma speed (km/s), and magnetic field strength \textbf{(G)} and the type II's speed obtained from the Richardson/Cane catalog \&\ the B field strength (G), which is obtained from the radio data analysis of type II burst as described in this section. 

\section{Results}
\label{section:sec3}
\subsection{Case study}
The dynamic spectra of the slowly drifting type II burst with a fundamental and harmonic band, which occurred on 6 Jul, 2006, at 08:23:12 UT, is displayed in Fig.\ref{fig:figure1}. The CME with a distinct three-part structure begins at 08:54 UT in the lower panel and is subsequently displayed, followed by a disturbance 
recorded near Earth at 21:36 UT on 9 Jul, 2006 \citep{Richardson2010}, Appendix \ref{tab:ICME}), where the ICME started on the 10 Jul, 2006 at 21:00 UT and ended on the 11 Jul, 2006 at 19:00 UT, lasting for 22 hours (see Fig.\ref{fig:figure2}). The CME propagated with a relatively higher speed of 911 km/s than the speed of a type II radio burst, estimated to be $\approx 369~km/s$ from the radio data analysis and the speed (km/s) of the associated ICME obtained from the Richardson catalog was 380 km/s. The difference in speeds of type II, CME, and ICME arises due to CMEs' interaction with the surroundings affects their speed  \citep{https://doi.org/10.1029/1999GL003639}; in this case, we have a CME with a fast speed of $\approx$ 1100 km/s \citep{
Kumari2023b}. The split-band type IIs are seen as well-accepted techniques to estimate magnetic field strengths just above the shock i.e. B in Gauss (G) derived from the type II radio bursts \citep{Smerd1975, Vrsnak2002, Cho2007, kumari2017c}, based on this and as described in section \ref{section:sec2}, we obtained 3 values for electron density (cm$^{-1}$), three values of each F$_{LB}$(MHz) and F$_{UB}$(MHz), three values of heliocentric distance (\( R_\odot \)) for each corresponding value i.e six values of heliocentric distance (\( R_\odot \)) and estimated three values of the magnetic field strength (B(G)). These values served as a basis for our power-law fit equation \citep{kumari2019direct}
\begin{equation}
\label{eq:eqq2}
    B =  2 \times r^{-a}  ~~[G]
\end{equation} 
where B is the average magnetic field, 
r is the heliocentric distance (in terms of \( R_\odot \), which is the radius of the photospheric Sun), 
At the range of 1.36\( R_\odot \) to 1.5\( R_\odot \) and lower band frequency range of F$_{LB}$ $\approx$ 141 MHz - 101 MHz, we estimated the average B field strength of 0.45 G using the equation \ref{eq:eqq2}, for the type II burst that occurred on 6 Jul, 2006.  The mean B(G) field of the associated with the upstream field just before the ICME ($B_{1AU} (G)$) was found to be 
$\mathbf{4.5 \times 10^{-5} (G)}$. This associated ICME lasted 22 hours (10 Jul, 2006 21:00 UT to 11 Jul, 2006 19:00 UT). The $B_{1AU} (G)$ was significantly attenuated as compared to estimated coronal B field strength (G) as expected due to the propagation of the CME in the heliosphere and its interaction with other magnetic structures there. Since the magnetic field derived from radio observations were from the lower frequency bands, which can be associated with the "undisturbed corona", and the ICME field is, by definition, the disturbed field, we used the B fields from upstream field (pre-shock region). 
In this study, we also explored the ICME parameters obtained from OMNIweb\footnote{\url{https://omniweb.gsfc.nasa.gov/index.html}} (NASA/GSFC), the average strength of the B field vector (G) was obtained to be $9.4 \times 10^{-5} (G)$ (mean) with the mean flow speed of 369 km/s for the start and end datetime of the ICME disturbance (see Figure \ref{fig:figure2}). These values are consistent with those of the values in the Richardson/Cane catalog.

\subsection{Statistical study of split-band type II bursts}

\begin{figure*}[ht!]
    \centering
    \includegraphics[width=18cm]{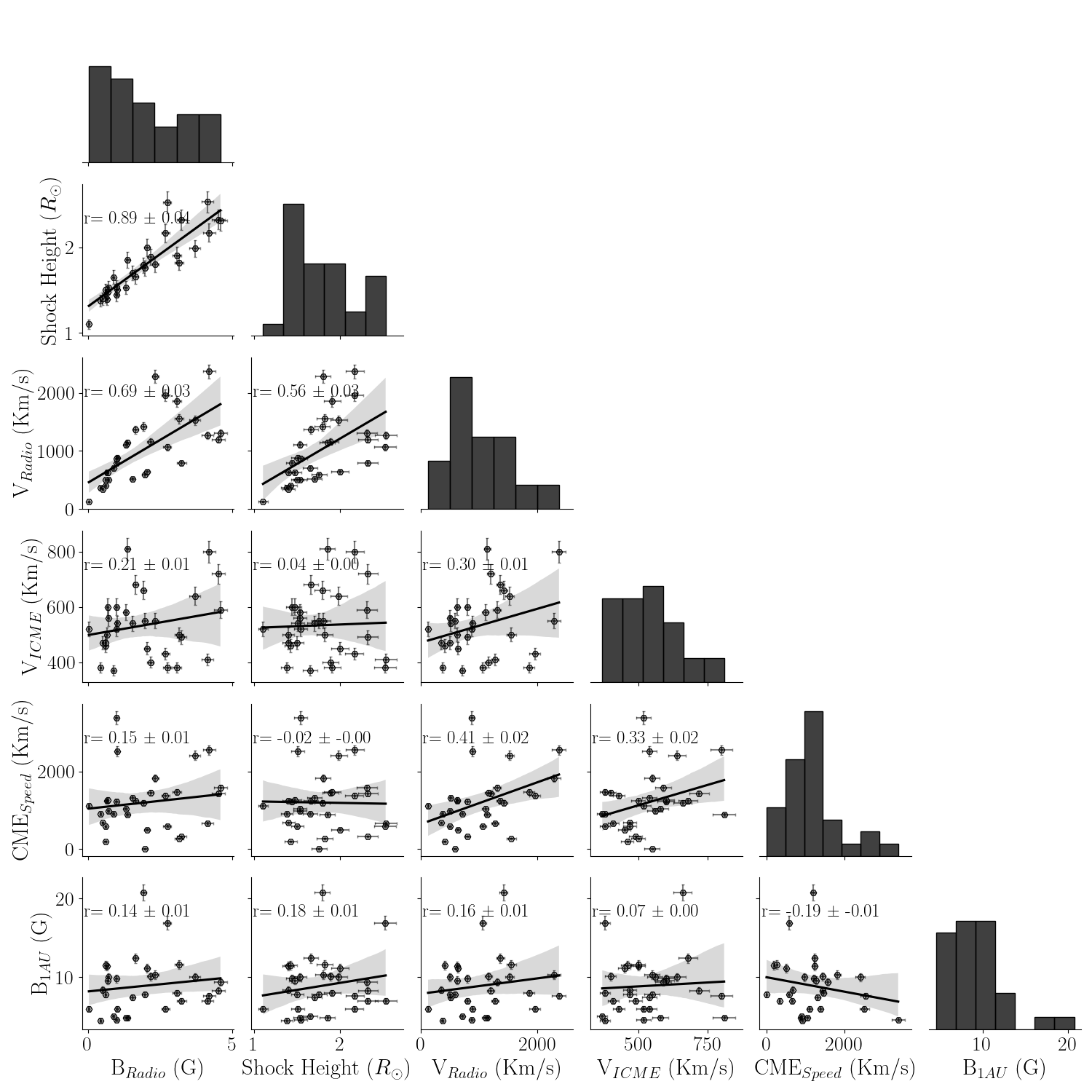}
    \caption{The correlation between various parameters of 31 CME-ICME-related events. The scattered hexagonal-shaped points are the data which contain shock height (R$_\odot$), F$_{start}$(MHz), F$_{end}$(MHz), Speed of Radio bursts(km/s), ICME speed(km/s), CME speed(km/s) and magnetic field strength derived from type IIs and obtained with ICMEs in-situ measurements.}
    \label{fig:figure3}
\end{figure*}

Similarly, we derived the B field strength values ($B_{Radio} (G)$) for 31 such split-band observations for the solar cycles of 23 and 24, varying from 0.04 G to 4.59 G for the heliocentric distance (r) ranging from 1.1\(R_\odot\) to 2.5\(R_\odot\). We note that the \textit{'a'} in the equation \ref{eq:eqq2} varies for individual split-band observations. We calculated the upstream magnetic field values, $B_{1AU} (G)$ at 1 AU ranging from $4.5 - 20.8 \times 10^{-5}~ G$ obtained from the OMNI database. A correlation between the near-sun coronal magnetic fields and near-earth upstream magnetic field strength observed just before the ICME was established by linearly fitting the magnetic fields ($B_{1AU} (G)$) at 1 AU and estimated ($B_{Radio} (G)$) derived from the Type II radio bursts) as seen in Fig. \ref{fig:figure3}, yielding a slope of 0.36 and an intercept of 8.19, suggesting linearly weak relation. The weak correlation was confirmed by the obtained correlation coefficient ($r$ of 0.14 (see Fig. \ref{fig:figure3})).  

We have used 4-fold Newkirk density model to estimate the shock heights (the radial distance from the Sun’s center to the leading edge of the shock wave that forms ahead of a CME as it propagates through the solar corona), in terms of $R_{\odot}$, close to the corona. We used 3-fold and 5-fold Newkirk density model to estimate the error bars for these shock height values (see Fig. \ref{fig:figure3}). These height estimate errors were then used to set the upper and lower limits of the error bars in the shock speed calculations. 
Similarly, we averaged over $\approx$ 15 min of upstream magnetic field values at 1 AU and used the rms values to estimate the uncertainties in $B_{1AU} (G)$ values in the pre-shock region (see Fig. \ref{fig:figure3}). The uncertainties in the speed of CME ($V_{CME}$) were estimated assuming linear propagation of the CME leading edge. The resulting uncertainties in the present study and correlation coefficients with error bars are shown in Fig. \ref{fig:figure3}. 

\section{Summary and Discussion}
\label{section:sec4}

Fig. \ref{fig:figure3} shows that the estimated magnetic fields near the Sun using radio techniques is not well correlated with the magnetic field measured at 1 AU using in-situ observations. Previous studies have suggested that suggest that metric radio observations can serve as effective proxies for estimating magnetic fields near the Sun \citep[][and the reference therein]{vrvsnak2008origin, kumari2017b, 2018JASTP.172...75M}. However to the best of our knowledge, there are not any event specific or comprehensive long-term studies using solar radio bursts to estimate the heliospheric magnetic fields. Our finding in this study (see Fig. \ref{fig:figure3} and Table \ref{tab:ICME}) strongly indicate that at least, no linear relation could be established using metric radio emissions to estimate the magnetic fields at 1 AU with acceptable error margins.
This suggests metric radio observations cannot serve as effective proxies for estimating magnetic fields close to the Earth. 
This could be due to the complex evolution of the magnetic field structures as it propagates through the heliosphere \citep[see for a review,][]{2013LRSP...10....5O}.
It is well known that CMEs are subjected to rotation, deflection, and interaction (interaction with solar wind/CME-CME interaction) \citep[][and the reference therein]{2021JGRA..12629770P, 2023ApJ...943L...8S}. However, we have not accounted for these conditions in our study due to the non-availability of radio data points between $\approx 2R_{\odot}$ and 1~AU, which are to be investigated in further studies/research with the availability of Parker Solar Probe (PSP) and Solar Orbiter (SolO) data. The CME-ICME speed is correlated, suggesting that other factors such as speed can help understand the predictive patterns. 
This study and the findings of this article can be summarized as follows:

   \begin{enumerate}
      \item Estimated B fields in the middle corona using split-band: We used the split-band observations to estimate magnetic fields within the middle corona. The split-band technique is an indirect measurement of magnetic field strengths just above the CME-driven shock by analyzing different frequency bands of type II radio bursts. 
      \item Statistical analysis of radio bursts associated with ICMEs: Through statistical analysis, we investigated radio bursts associated with ICMEs. We also examined the start and end frequencies, bandwidth, and duration of radio emissions observed during ICME events. By this, we aimed to identify a relation indicative of magnetic field variations and associated CMEs and ICMEs.
      \item Relation between the magnetic field near the Sun and at 1 AU: Our results show almost no clear linear relation between the magnetic field strength near the Sun obtained with metric radio observations and in-situ measurements of the magnetic field associated with upstream/shock transition field just before the ICME. Hence, we could not establish any empirical relation that provides a means to estimate the magnetic field intensity when the disturbance reaches the Earth, based on measurements obtained closer to the Sun. This indicates that metric radio observations are become less reliable for estimating magnetic fields closer to Earth.
   \end{enumerate}

\begin{acknowledgements}
{All the authors acknowledge the event lists provided by the Space Weather Prediction Center and the ICME catalog compiled by Ian Richardson and Hilary Cane. The authors acknowledge the sunpy package, the Coordinated Data Analysis Workshop (CDAW) catalog, Solar Electro-Optical Network's (SEON) Radio Spectral Telescope Network (RSTN) data, eCallisto data, National Institute of Information and Communications Technology's (NICT) Hiraiso Radio Spectrograph (HiRAS) data, Solar and Heliospheric Observatory's (SOHO) Large Angle and Spectrometric Coronagraph (LASCO), and NASA/GSFC's OMNIweb database.}
\end{acknowledgements}

\begin{landscape}
\begin{table}[ht!]

\section{Appendix}

\begin{tabular}{llllllllllll}
\hline
         &       &       & Type II &           &     &     &  & CME   &     &      \\ \hline
Date     & Time  &       &      &   & Frequency &     &     &  &   &     &      \\ 
 &
  \begin{tabular}[c]{@{}l@{}}T$_{start}$\\ (UT)\end{tabular} &
  \begin{tabular}[c]{@{}l@{}}T$_{end}$\\ (UT)\end{tabular} &
  \begin{tabular}[c]{@{}l@{}}T$_{duration}$\\ (min)\end{tabular} &
  \begin{tabular}[c]{@{}l@{}}Shock Height\\($R_{\odot}$)\end{tabular} &
  \begin{tabular}[c]{@{}l@{}}F$_{start}$\\ (MHz)\end{tabular} &
  \begin{tabular}[c]{@{}l@{}}F$_{end}$\\ (MHz)\end{tabular} &
  \begin{tabular}[c]{@{}l@{}}BW\\ (MHz)\end{tabular} &
  \begin{tabular}[c]{@{}l@{}}T$_{start}$\\ (UT)\end{tabular} &
  \begin{tabular}[c]{@{}l@{}}Width\\ (deg)\end{tabular} &
  \begin{tabular}[c]{@{}l@{}}v$_{CME}$\\ (km/s)\end{tabular} \\ \hline
  \\
12 Feb, 2000 & 04:06 & 04:17 & 11    &1.10 & 80   & 30   & 50 & 04:31 & 360 & 1107 \\
16 Sep, 2000 & 04:17 & 04:33 & 16    &1.44 & 180  & 25    & 155 & 05:18 & 360 & 1215 \\
25 Nov, 2000 & 01:07 & 01:16 & 09    &1.50 & 147  & 25    & 122 & 01:31 & 360 & 2519 \\
10 Apr, 2001 & 05:13 & 05:17 & 04    &1.98 & 80   & 30   & 50 & 05:30 & 360 & 2411 \\
09 Oct, 2001 & 10:54 & 11:02 & 08    &1.53 & 180  & 25    & 155 & 11:30 & 360 & 973 \\
17 Nov, 2001 & 04:50 & 04:55 & 05    &2.16 & 145  & 45    & 100 & 05:30 & 360 & 1379 \\
22 Nov, 2001 & 22:31 & 22:41 & 10    &2.32 & 116  & 25    & 91 & 23:30 & 360 & 1437 \\
26 Dec, 2001 & 05:02 & 05:19 & 17    &1.88 & 180  & 25    & 155 & 05:30 & 212 & 1446 \\
17 Apr, 2002 & 08:08 & 08:23 & 15    &1.39 & 79   & 25   & 54 & 08:26 & 360 & 1240 \\
29 Jul, 2002 & 02:40 & 02:43 & 03    &1.42 & 270  & 65    & 205 & 12:07 & 095  & 190 \\
29 May, 2003 & 01:06 & 01:11 & 05    &1.66 & 80   & 30   & 50 & 01:27 & 360 & 1237 \\
12 Sep, 2004 & 01:41 & 01:50 & 09    &1.76 & 180  & 57    & 123 & 00:36 & 360 & 1328 \\
10 Nov, 2004 & 02:07 & 02:15 & 08    &1.54 & 260  & 70    & 190 & 02:26 & 360 & 3387 \\
17 Jan, 2005 & 09:44 & 09:47 & 03    &2.17 & 65   & 25   & 40 & 09:30 & 360 & 2547 \\
20 Jan, 2005 & 06:44 & 07:00 & 16    &1.85 & 180 & 160    & 20 & 06:54 & 360 & 882 \\
22 Aug, 2005 & 01:02 & 01:08 & 06    &1.79 & 60   & 30   & 30 & 01:31 & 360 & 1194 \\
06 Jul, 2006 & 08:24 & 08:41 & 17    &1.38 & 180  & 25    & 155 & 08:54 & 360 & 911 \\
14 Dec, 2006 & 22:09 & 22:13 & 04    &1.53 & 180  & 25    & 155 & 22:30 & 360 & 1042 \\
15 Feb, 2011 & 01:52 & 02:00 & 08    &1.40 & 260  & 57    & 203 & 02:36 & 360 & 669 \\
04 Aug, 2011 & 03:54 & 04:03 & 09    &1.70 & 180  & 66    & 114 & 04:12 & 360 & 1315 \\
06 Sep, 2011 & 22:19 & 22:34 & 15    &1.50 & 300  & 40    & 260 & 23:05 & 360 & 575 \\
09, Nov, 2011 & 13:11 & 13:31 & 20    &1.65 & 180  & 30    & 150 & 13:30 & 360 & 907 \\
19 Jan, 2012 & 12:52 & 13:01 & 09    &2.00 & 084  & 29    & 055 & 14:36 & 065  & 498 \\
07 Mar, 2012 & 01:09 & 01:29 & 20    &1.80 & 180  & 25    & 155 & 00:24 & 360 & 1825 \\
04 Jul, 2012 & 16:42 & 17:04 & 22    &2.53 & 181  & 25    & 156 & 17:24 & 360 & 662 \\
12 Jul, 2012 & 16:25 & 16:53 & 28    &2.32 & 82   & 25   & 57 & 16:48 & 025  & 329 \\
02 Apr, 2014 & 13:23 & 13:25 & 02    &1.91 & 121  & 28    & 93 & 13:48 & 360 & 1471 \\
10 Sep, 2014 & 17:27 & 17:57 & 30    &1.48 & 180  & 25    & 155 & 18:00 & 360 & 1267 \\
17 Dec, 2014 & 04:44 & 05:01 & 27    &2.52 & 70   & 23   & 47 & 05:00 & 360 & 587 \\
04 Nov, 2015 & 03:23 & 03:34 & 11    &1.82 & 180  & 41    & 139 & 14:48 & 064  & 272 \\
06 Sep, 2017 & 12:02 & 12:21 & 19    &2.31 & 081  & 25    & 56 & 12:24 & 360 & 1571 \\ \hline
\end{tabular}
\caption{The table contains various parameters of type II bursts and CMEs, that are associated with the ICMEs in table \ref{tab:ICME}}
\label{tab:CME+typeII}
\end{table}
\end{landscape}

\begin{landscape}
\begin{table}[ht!]
\begin{tabular}{llclclclccccllc}
\hline
 &
   &
  \multicolumn{1}{l}{} &
   &
  \multicolumn{1}{l}{ICME} &
   &
  \multicolumn{1}{l}{} &
   &
  \multicolumn{1}{l}{} &
  \multicolumn{1}{l}{} &
  \multicolumn{1}{l}{} &
  \multicolumn{1}{l}{} &
   &
   \\ \hline
 &
  Disturbance &
  \multicolumn{1}{l}{} &
   &
  \multicolumn{1}{l}{ICME Plasma/Field} &
   &
  \multicolumn{1}{l}{} &
   &
  \multicolumn{1}{l}{} &
  \multicolumn{1}{l}{} &
  \multicolumn{1}{l}{} &
  \multicolumn{1}{l}{} &
   &
   \\
Date &
  \begin{tabular}[c]{@{}l@{}}Time\\ T$_s$\\ (UT)\end{tabular} &
   &
  \multicolumn{1}{l}{\begin{tabular}[c]{@{}l@{}}Start Date\\ \end{tabular}} &
  \begin{tabular}[c]{@{}l@{}}T$_{start}$\\ (UT)\end{tabular} &
  \multicolumn{1}{l}{\begin{tabular}[c]{@{}l@{}}End Date\\ \end{tabular}} &
  \begin{tabular}[c]{@{}l@{}}T$_{end}$\\ (UT)\end{tabular} &
  \multicolumn{1}{l}{\begin{tabular}[c]{@{}l@{}}ICME$_{\text{plasma Speed}}$ \\ (km/s)\end{tabular}} &
  \multicolumn{1}{l}{\begin{tabular}[c]{@{}l@{}}v$_{max}$\\ (km/s)\end{tabular}}  &
  \multicolumn{1}{l}{\begin{tabular}[c]{@{}l@{}}\textbf{${B_{1AU}}$}\\ $(\times 10^{-5} (G))$\end{tabular}} &
   &
  \begin{tabular}[c]{@{}l@{}}v$_{Radio}$\\ (km/s)\end{tabular} &
  \begin{tabular}[c]{@{}l@{}}B$_{Radio}$\\ (G)\end{tabular} \\ \hline
  \\
  14 Feb, 2000 &
  07:31  &
   &
  14 Feb, 2000 &
  12:00 &
  16 Feb, 2000 &
  08:00 &
  520 &
  680 &
  6 &
   &
  122 &
  0.04\\
17 Sep, 2000 &
  16:57  &
   &
  17 Sep, 2000 &
  21:00 &
  21 Sep, 2000 &
  00:00 &
  600 &
  840 &
  9.8 &
   &
  793 &
  0.99\\
28 Nov, 2001 &
  05:30  &
   &
  28 Nov, 2001 &
  11:00 &
  29 Nov, 2001 &
  22:00 &
  540 &
  580 &
  6&
   &
  883 &
  1.01\\
11 Apr, 2001 &
  13:43  &
   &
  11 Apr, 2001 &
  22:00 &
  13 Apr, 2001 &
  07:00 &
  640 &
  740 &
  10 &
   &
  1525 &
  3.71\\
11 Oct, 2001 &
  17:01  &
   &
  12 Oct, 2001 &
  04:00 &
  12 Oct, 2001 &
  09:00 &
  560 &
  570 &
  10 &
   &
  503 &
  0.71\\
19 Nov, 2001 &
  18:15  &
   &
  19 Nov, 2001 &
  22:00 &
  21 Nov, 2001 &
  13:00 &
  430 &
  570 &
  6 &
   &
  1953 &
  2.69\\
  24 Nov, 2001 &
  06:56  &
   &
  24 Nov, 2001 &
  14:00 &
  25 Nov, 2001 &
  20:00 &
  720 &
  1040 &
  8.3 &
   &
  1196 &
  4.52\\
29 Dec, 2001 &
  05:38  &
   &
  30 Dec, 2001 &
  00:00 &
  30 Dec, 2001 &
  18:00 &
  400 &
  460 &
  10.1 &
   &
  1158 &
  2.17\\
19 Apr, 2002 &
  08:35  &
   &
  20 Apr, 2002 &
  00:00 &
  21 Apr, 2002 &
  18:00 &
  500 &
  640 &
  11.4 &
   &
  0624 &
  0.65\\
01 Aug, 2002 &
  23:09 &
   &
  02 Aug, 2002 &
  06:00 &
  04 Aug, 2002 &
  02:00 &
  460 &
  520 &
  11.5 &
   &
  404 &
  0.60\\
30 May, 2003 &
  16:00 &
   &
 30 May, 2003 &
  22:00 &
01 Jun, 2003 &
  01:00 &
  680 &
  780 &
  12.4 &
   &
  1364 &
  1.66\\
13 Sep, 2004 &
  20:03 &
   &
  14 Sep, 2004 &
  15:00 &
 16 Sep, 2004 &
  12:00 &
  550 &
  600 &
  7.8 &
   &
  586 &
  1.97\\
11 Nov, 2004 &
  17:10 &
   &
 12 Nov, 2004 &
  08:00 &
13 Nov, 2004 &
  23:00 &
  520 &
  670 &
  4.6 &
   &
  869 &
  0.99\\
  18 Jan, 2005 &
  21:00 &
   &
  18 Jan, 2005 &
  23:00 &
  20 Jan, 2005 &
  03:00 &
  800 &
  960 &
  7.6 &
   &
  2369 &
  4.20\\
21 Jan, 2005 &
  17:11 &
   &
  21 Jan, 2005 &
  19:00 &
  22 Jan, 2005 &
  17:00 &
  810 &
  960 &
  4.8 &
   &
  1138 &
  1.38\\
24 Aug, 2005 &
  06:13 &
   &
  24 Aug, 2005 &
  14:00 &
  24 Aug, 2005 &
  23:00 &
  660 &
  710 &
  20.8 &
   &
  1417 &
  1.93\\
09 Jul, 2006 &
  21:36 &
   &
  10 Jul, 2006 &
  21:00 &
  11 Jul, 2006 &
  19:00 &
  380 &
  430 &

  4.5&
   &
  369 &
  0.45\\
16 Dec, 2006 &
  17:55 &
   &
  17 Dec, 2006 &
  00:00 &
  17 Dec, 2006 &
  17:00 &
  580 &
  680 &
  4.8 &
   &
  1106 &
  1.31\\
18 Feb, 2011 &
  01:30 &
   &
  18 Feb, 2011 &
  19:00 &
  20 Feb, 2011 &
  08:00 &
  470 &
  600 &
  8.4 &
   &
  346 &
  0.52\\
05 Aug, 2011 &
  17:51 &
   &
  06 Aug, 2011 &
  22:00 &
  07 Aug, 2011 &
  22:00 &
  540 &
  610 &
  7.4 &
   &
  513 &
  1.54\\
09 Sep, 2011 &
  12:42 &
   &
  10 Sep, 2011 &
  03:00 &
  10 Sep, 2011 &
  15:00 &
  470 &
  530 &
  7.8 &
   &
  501 &
  0.61\\
12 Nov 2011 &
  05:59 &
   &
  13 Nov 2011 &
  10:00 &
  15 Nov 2011 &
  02:00 &
  370 &
  460 &
  5 &
   &
  703 &
  0.88\\
22 Jan, 2012 &
  06:11 &
   &
  22 Jan, 2012 &
  23:00 &
  23 Jan, 2012 &
  07:00 &
  450 &
  460 &
  11.1 &
   &
  640 &
  2.05\\
08 Mar, 2012 &
  11:03 &
   &
  09 Mar, 2012 &
  03:00 &
  11 Mar, 2012 &
  07:00 &
  550 &
  890 &
  10.3 &
   &
  2285 &
  2.33\\
08 Jul, 2012 &
  08:00 &
   &
  09 Jul, 2012 &
  00:00 &
  09 Jul, 2012 &
  14:00 &
  410 &
  450 &
  7 &
   &
  1269 &
  4.14\\
14 Jul, 2012 &
  18:09 &
   &
  15 Jul, 2012 &
  06:00 &
  17 Jul, 2012 &
  05:00 &
  490 &
  670 &
  7 &
   &
  794 &
  3.25\\
  05 Apr, 2014 &
  10:00 &
   &
  05 Apr, 2014 &
  22:00 &
  07 Apr, 2014 &
  05:00 &
  380 &
  500 &
  8 &
   &
  1853 &
  3.09\\
12 Sep, 2014 &
  15:53 &
   &
  12 Sep, 2014 &
  22:00 &
  14 Sep, 2014 &
  02:00 &
  600 &
  720 &
  9.6 &
   &
  626 &
  0.68\\
  21 Dec, 2014 &
  19:11 &
   &
  22 Dec, 2014 &
  04:00 &
  22 Dec, 2014 &
  17:00 &
  380 &
  430 &

  17 &
   &
  1063 &
  2.74\\
06 Nov, 2015 &
  1818 &
   &
  07 Nov, 2015 &
  06:00 &
  08 Nov, 2015 &
  16:00 &
  500 &
  680 &

  11.6 &
   &
  1550 &
  3.16\\
07 Sep, 2017 &
  2302 &
   &
  08 Sep, 2017 &
  11:00 &
  10 Sep, 2017 &
  21:00 &
  590 &
  800 &
  9.4 &
   &
  1303 &
  4.59\\
 &
   &
  \multicolumn{1}{l}{} &
   &
  \multicolumn{1}{l}{} &
   &
  \multicolumn{1}{l}{} &
   &
  \multicolumn{1}{l}{} &
  \multicolumn{1}{l}{} &
  \multicolumn{1}{l}{} &
  \multicolumn{1}{l}{} &
   &
   \\
 &
   &
  \multicolumn{1}{l}{} &
   &
  \multicolumn{1}{l}{} &
   &
  \multicolumn{1}{l}{} &
   &
  \multicolumn{1}{l}{} &
  \multicolumn{1}{l}{} &
  \multicolumn{1}{l}{} &
  \multicolumn{1}{l}{} &
   &
   \\
 &
   &
  \multicolumn{1}{l}{} &
   &
  \multicolumn{1}{l}{} &
   &
  \multicolumn{1}{l}{} &
   &
  \multicolumn{1}{l}{} &
  \multicolumn{1}{l}{} &
  \multicolumn{1}{l}{} &
  \multicolumn{1}{l}{} &
   &
   \\ \hline
\end{tabular}
\caption{The table contains various ICME parameters associated with the type II bursts and CMEs in table \ref{tab:CME+typeII}}
\label{tab:ICME}
\end{table}
\end{landscape}

\end{document}